\documentclass[aps,prl,twocolumn,superscriptaddress,nofootinbib]{revtex4-2}
\usepackage{amsmath,amssymb,bm,mathtools}
\usepackage[colorlinks=true,linkcolor=blue,citecolor=blue,urlcolor=blue]{hyperref}

\begin{document}

\title{Resolving the Conformal Mode Problem with Einstein-Cartan Gravity}

\author{Stephon Alexander}
\email{stephon\_alexander@brown.edu}
 \affiliation{BCTPI, Department of Physics, Brown University, Providence, Rhode Island 02912, USA}

\author{Raymond Isichei}
\email{ri221@ic.ac.uk}
\affiliation{Abdus Salam Centre for Theoretical Physics, Imperial College London, Prince Consort Road, London SW7 2BZ, United Kingdom}

\author{Jo\~ao Magueijo}
\email{j.magueijo@imperial.ac.uk}
\affiliation{Abdus Salam Centre for Theoretical Physics, Imperial College London, Prince Consort Road, London SW7 2BZ, United Kingdom}

\begin{abstract}
 We show that, when $\Lambda>0$, the conformal factor problem is absent in first order Einstein-Cartan gravity in which the spin connection and tetrad are independent off-shell. At fixed connection, a conformal rescaling of the tetrad produces no conformal kinetic term so that the action is bounded from below. The original unbounded Einstein-Hilbert action, the Einstein-Hilbert action without the problematic conformal kinetic term and conformally invariant $R^{2}/\Lambda$ gravity arise as classical saddles of the Einstein-Cartan action with respect to the connection and conformal mode of the tetrad.\\
\end{abstract}

\maketitle

\paragraph{Introduction.}
The success of Euclidean methods in quantum field theory, including Yang-Mills theory, motivated
their extension to the quantization of gravity. An early success was
the semiclassical derivation of black hole thermodynamics from the
Euclidean gravitational action \cite{GH1977}. The extension to an
integral over fluctuating geometries exposed a difficulty that has
persisted for nearly five decades. In 1978, Gibbons, Hawking, and Perry
showed that the Euclidean Einstein action is unbounded below along
the conformal direction and proposed rotating the conformal factor
into the complex plane \cite{GHP1978,GP1978}. Subsequent work has
related this prescription to the gauge-fixed measure and the
nonpropagating character of the conformal mode
\cite{Schleich1987,MazurMottola1990,DasguptaLoll2001,Hartle}.

The origin of the problem is apparent from the Euclidean
Einstein--Hilbert action
\begin{equation}
    S_{\rm EH,E}[g]
    =-\frac{1}{16\pi G}\int d^4x\,\sqrt{g}\,
      \bigl(R-2\Lambda\bigr).
    \label{EH}
\end{equation}
Under a local conformal rescaling
$g_{\mu\nu}\mapsto e^{2\phi}g_{\mu\nu}$, the action becomes, up to a
boundary term,
\begin{equation}
\begin{aligned}
    S_{\rm EH,E}^{\rm conf}[g,\phi]
    =-\frac{1}{16\pi G}\int d^4x\,\sqrt{g}\,
    \bigl[&e^{2\phi}R-2\Lambda e^{4\phi}\\
           &+6e^{2\phi}(\nabla\phi)^2\bigr].
\end{aligned}
\label{confEH}
\end{equation}
The conformal gradient term enters the action with a negative sign.
Conformal fluctuations of bounded amplitude and arbitrarily short
wavelength therefore drive the action to $-\infty$, even for
$\Lambda>0$, and the Euclidean weight enhances them. How does this
instability arise when the geometry and the connection are varied
independently?

Einstein--Cartan gravity provides a natural setting in which to
address this question. Its fundamental fields are the tetrad and spin
connection, and these are independent off shell, before the equations
of motion are imposed \cite{Kibble1961,Hehl1976}. The usual derivation
of the conformal instability relies on the fact that the
Levi--Civita connection is built from derivatives of the metric.
Once the metric is rescaled, derivatives of the conformal factor enter
the connection and therefore the curvature. In the first-order theory
this implication is absent off shell. A local rescaling of the tetrad
does not require a simultaneous change of the independent spin
connection.

In this Letter we show that, at fixed spin connection, the conformal
factor problem in its historical form does not arise. Conformal
rescalings of the tetrad yield no derivatives of the conformal factor
in the Euclidean Einstein--Cartan action, which is bounded below
along this direction for $\Lambda>0$. The negative conformal kinetic
term reappears when the connection is taken to its torsion-free
stationary value for the rescaled tetrad. If the total spin connection
is held fixed, the change in contorsion compensates for the change in
the Levi--Civita connection, and the unstable kinetic term is exactly cancelled. On the
positive-curvature branch, the elimination of the conformal factor at its
algebraic saddle yields a locally Weyl invariant action proportional
to $R_\omega^2/\Lambda$, where $R_\omega$ is the curvature scalar of
the independent connection. The first-order formulation thus
identifies which coupled variations of the tetrad and spin connection
recover the familiar instability and opens a route toward resolving
the conformal factor problem through a joint integration contour.

% Add to the bibliography:
% \bibitem{GH1977}
% G.~W.~Gibbons and S.~W.~Hawking,
% Phys. Rev. D \textbf{15}, 2752 (1977).

%The remaining contour problem belongs to the connection, whose integration cycle we argue is inherited from Lorentzian theory. We illustrate this mechanism in closed de Sitter minisuperspace, where the scale variable is never rotated and the connection contour yields the Chern–Simons–Kodama state and its Airy transform, the Hartle–Hawking wave function \cite{Kodama1990,SmolinSoo1995,HH1983,Magueijo2020,IM}. 

\begin{center}
    {\it Einstein-Cartan $\rightarrow$ Einstein-Hilbert}
\end{center}
In $d=4$ Euclidean Einstein-Cartan formalism, metric information is contained in the tetrad 1-form $e^{I}=e^{I}_{\mu}dx^{\mu}$ which obeys the relation $g_{\mu \nu}dx^{\mu}dx^{\nu}=\delta_{IJ}e^{I}e^{J}$ where $\delta_{IJ}={\rm diag}(1,1,1,1)$ is an internal Euclidean metric. Affine information is encoded in the spin connection 1-form $\omega^{IJ}=\omega^{IJ}_{\mu}dx^{\mu}$ associated with internal $\rm SO(4)$ gauge transformations and related to the usual Christoffel symbols $\Gamma^{\lambda}_{\mu \nu}$ via the tetrad postulate $\partial_{\nu}e^{I}_{\mu}+\omega^{I}{}_{J\nu}e^{J}_{\mu}=\Gamma^{\lambda}_{\mu \nu}e^{I}_{\lambda}$. Defining the covariant derivative operation as $D_{\omega}(*)=d(*)+\omega\wedge(*)$, torsion $T^{I}$ is defined as
\begin{equation}
    T^{I}:=D_{\omega}e^{I}=de^{I}+\omega^{I}{}_{J}\wedge e^{J}.
    \label{torsion00}
\end{equation}
The spin connection $\omega^{IJ}$ may be further decomposed as $\omega^{IJ}=\omega^{IJ}_{(LC)}+K^{IJ}$ where $\omega^{IJ}_{(LC)}$ is the torsion-free Levi-Civita connection obeying
\begin{equation}
    de^{I}+\omega^{I}{}_{J(LC)}\wedge e^{J}=0,
    \label{torsion01}
\end{equation}
equivalent to the tetrad postulate. Granting (\ref{torsion00}) and (\ref{torsion01}) implies that $T^{I}=K^{I}{}_{J}\wedge e^{J}$ so that torsion is entirely determine by the antisymmetric contorsion 1-form $K^{IJ}$ \cite{Hehl1976, Hehl1995, Ham, Greek}. From $\omega^{IJ}$, the curvature 2-form can be derived as
\begin{equation}
\begin{gathered}
        D_{\omega}\omega^{IJ}:=R^{IJ}[\omega],\\
        R^{IJ}[\omega]=d\omega^{IJ}+\omega^{I}{}_{M}\wedge\omega^{MJ}.
        \label{curv}
\end{gathered}
\end{equation}
A consequence of the decomposition $\omega^{IJ}=\omega^{IJ}_{(LC)}+K^{IJ}$ is that (\ref{curv}) may be decomposed as the sum of Levi-Civita connection curvature and contorsion curvature as
\begin{equation}
    R^{IJ}[\omega]=R^{IJ}\big[\omega_{(LC)}\big]+D_{\omega}K^{IJ} +K^{I}{}_{M}\wedge K^{MJ}.
    \label{curv1}
\end{equation}
The gravitational action defined by the independent fields $e^{I}$ and $\omega^{IJ}$ which contains $R^{IJ}[\omega]$, is the Euclidean Einstein-Cartan action
\begin{align}
\begin{split}
        &S_{\rm EC, E}[e,\omega]=\\
        -&\frac{1}{32\pi G}\int \epsilon_{IJKL} \ e^{I}\wedge e^{J} \wedge\bigg[R^{KL}[\omega]-\frac{\Lambda}{6}e^{K}\wedge e^{L} \bigg] .
    \label{ECE}
\end{split}
\end{align}
The variation of (\ref{ECE}) with respect to the connection $\omega$ yields
\begin{equation}
D_\omega(e^I\wedge e^J)=0.
\label{torsioneq}
\end{equation}
For $e^{I}\neq0$, (\ref{torsioneq}) enforces $T^{I}=K^{I}{}_{J}\wedge e^{J}=0$ so that (\ref{torsion01}) is true and $\omega^{IJ}=\omega^{IJ}_{(LC)}$ on-shell. This, in turn, implies that (\ref{curv1}) simplifies to 
\begin{equation}
R^{IJ}[\omega]=R^{IJ}\big[\omega _{(LC)}\big]
\label{curveq}
\end{equation}
on-shell. The perspective we espouse in this letter is that the {\it on-shell} Einstein-Cartan action $S_{\rm EC,E}\big[e,\omega_{(LC)}\big]$ is given by
\begin{align}
\begin{split}
    &S_{\rm EC,E}\big[e,\omega_{(LC)}\big]=\\
    -&\frac{1}{32\pi G}\int \epsilon_{IJKL} \ e^{I}\wedge e^{J}\wedge\bigg[R^{KL}\big[\omega_{(LC)}\big]-\frac{\Lambda}{6}e^{K}\wedge e^{L} \bigg].
\end{split}
\end{align}
Using the relevant $\epsilon-\delta$ identities, the Euclidean Einstein-Hilbert action can be recovered so that standard GR is a connection saddle-point of Einstein-Cartan gravity \cite{Hehl1976, Daum, Wheel, Belt, Pekar, Shapiro}. 

%In the Supplemental material we use mini-superspace FRW geometry as an illustrative example of this rationale.

\begin{center}
    {\it Conformal Transformations of the Euclidean Einstein-Cartan action}
\end{center}
Given the off-shell independence of $e^{I}$ and $\omega^{IJ}$ in addition to the definition $g_{\mu\nu}dx^{\mu}dx^{\nu}=\delta_{IJ}e^{I}e^{J}$, the conformal rescaling of the metric $g_{\mu \nu}\rightarrow e^{2\phi}g_{\mu \nu}$ can be translated into the rescalings \cite{Karahan, Shap, Chakra, Obu, Shapiro, Shapiro2, Chile, N1}
\begin{equation}
e^I\longrightarrow e^{\phi}e^I,
\qquad
\omega^{IJ}\longrightarrow\omega^{IJ}.
\label{weyl}
\end{equation}
As $R^{IJ}$ depends only on the independent connection $\omega^{IJ}$ which doesn't scale at the level of the action, the conformally transformed action is
\begin{align}
    \begin{split}
        &S_{\rm EC,E}^{\rm conf}[e,\omega, \phi]=\\
       -& \frac{1}{32\pi G}\int \epsilon_{IJKL} \ e^{I}\wedge e^{J} \wedge \bigg[e^{2\phi}R^{KL}[\omega]-e^{4\phi}\frac{\Lambda}{6}e^{K}\wedge e^{L} \bigg].
    \end{split}
    \label{confEC}
\end{align}
Notice that $R^{KL}$ and $\Lambda$ in (\ref{confEC}) are multiplied by the same conformal prefactors as the $R$ and $\Lambda$ terms in the transformed Einstein-Hilbert action (\ref{confEH}). However, there is one critical difference, in (\ref{confEC}) there are no derivatives of $\phi$. The Lagrangian is a pure potential of the form $\mathcal{L}_{ \rm EC, E}^{\rm conf}=-Ay+B y^{2}/6$, where $y=e^{2\phi}$. The action (\ref{confEC}) is therefore bounded from below along the conformal direction if there exists a $y_{\rm min}=e^{2\phi_{\rm min}}>0$. The conformal mode saddle point of (\ref{confEC}) is
\begin{equation}
    y_{\rm min}=e^{2\phi_{\rm min}}=\frac{3R}{\Lambda}
    \label{phimin}
\end{equation}
  so that $y_{\rm min}>0$ when both $R>0$ and $\Lambda>0$, which corresponds to classical de Sitter geometry. Consequently, the on-shell action evaluated on the conformal mode saddle is
 \begin{align}
 \begin{split}
     &S_{\rm EC, E}[e, \omega,\phi_{\rm min}]=\\
     -&\frac{3}{64\pi G \Lambda}\int \epsilon_{IJKL} \  e^{I}\wedge e^{J}\wedge R^{KL}[\omega]\times R[\omega],
     \label{phishell}
 \end{split}
 \end{align}
where $R[\omega]=e^{\mu}_{M}e^{\nu}_{N}R^{MN}_{\mu \nu}[\omega]$ and $1/G\Lambda$ is dimensionless. Naturally, the elimination of the conformal mode leads to (\ref{phishell}) being invariant under conformal transformations as $e^{I}\wedge e^{J}\rightarrow e^{2\phi} e^{I}\wedge e^{J}$, $R^{KL}\rightarrow R^{KL}$ and $R\rightarrow e^{-2\phi}R$.

\begin{center}
    \textit{Resolution of the conformal factor problem}
\end{center}

In what follows, we show that for \(\Lambda>0\) the Euclidean
Einstein--Cartan action is bounded below under local conformal
rescalings of the tetrad at fixed total spin connection. The
contorsion shift compensates for the change in the Levi--Civita
connection, leaving the full curvature unchanged and the action
free of a conformal kinetic term.

The boundedness of the action (\ref{confEC}) from below by its minimum (\ref{phishell}) for $\Lambda>0$ was deduced without reference to a problematic conformal kinetic term. This is primarily due to the analysis being off-shell with respect to $\omega^{IJ}$.\\
In the first-order formulation, the tetrad and the full spin
connection are independent variables. We consider a local
conformal rescaling \(e'^I=e^\phi e^I\) with invariant full
connection, \(\omega^{\prime IJ}=\omega^{IJ}\).  Using (\ref{torsion01}), the Levi-Civita connection after a conformal transformation $\omega^{\prime IJ}_{(LC)}(e^{\prime})$ can be expressed in terms of the original Levi-Civita connection $\omega^{ IJ}_{(LC)}(e)$ as
\begin{equation}
    \omega_{(LC)}'{}^I{}_J\wedge e^J
    =\omega_{(LC)}^I{}_J\wedge e^J-d\phi\wedge e^I .
\end{equation}
It follows that \cite{Shapiro,Shapiro2,Chile}
\begin{equation}
\omega_{(LC)}^{\prime IJ}=\omega^{IJ}_{(LC)}+\theta^{IJ},
    \qquad
    \theta^{IJ}=e^I\nabla^J\phi-e^J\nabla^I\phi ,
    \label{omega02}
\end{equation}
where \(\nabla^I\phi\) is the frame derivative defined by \(e^I\).
To see how the problematic conformal kinetic term re-emerges on-shell and how it's cancelled in Einstein-Cartan formalism, we rely on the invariance of the full connection $\omega^{\prime IJ}=\omega^{IJ}$. Expressing this equation as $\omega_{(LC)}^{\prime IJ}+K^{\prime IJ}=\omega_{(LC)}^{IJ}+K^{IJ}$ implies that, upon the rescaling (\ref{omega02}), the contorsion should transform as \cite{Shapiro,Shapiro2,Chile}
\begin{equation}
    K^{\prime IJ}=K^{IJ}-\theta^{IJ} \ .
    \label{conto02}
\end{equation}
The equations (\ref{omega02}) and (\ref{conto02}) hold for arbitrary contorsion $K^{IJ}$. Therefore, the shift in contorsion compensates for the shift in the
Levi-Civita connection, leaving the full curvature two-form
invariant, or
\begin{equation}
    R^{IJ}[\omega']=R^{IJ}[\omega] \ .
    \label{curv03}
\end{equation}
To resolve the conformal factor problem by cancelling the conformal kinetic term, we set the original pair $(e,\omega)$ to be torsion free so that the inital full connection is $\omega^{IJ}=\omega^{IJ}_{(LC)}$ with the initial contorsion vanishing ($K^{IJ}=0$). The transformation (\ref{conto02}) simplifies to $K^{\prime IJ}=-\theta^{IJ}$. Thus, in (\ref{curv03}), $R^{IJ}[\omega]=R^{IJ}[\omega_{(LC)}]$. Note that the curvature difference induced by (\ref{omega02}) is
\begin{equation}
    R^{IJ}\big[\omega^{\prime}_{(LC)}\big]-R^{IJ}\big[\omega_{(LC)}\big]=D_{\omega}\theta^{IJ}+\theta^{IM}\wedge \theta^{J}{}_{M} \ ,
\end{equation}
while the difference in the full curvatures is
\begin{equation}
    R^{IJ}[\omega^{\prime}]-R^{IJ}[\omega]=D_{\omega}K^{\prime IJ}-K^{\prime I}{}_{M}\wedge K^{\prime MJ} \ .
\end{equation}
Since the full connection doesn't change, it follows that the curvatures don't changes so that (\ref{curv03}) becomes

    \begin{equation}
    R^{IJ}\big[\omega^{\prime}_{(LC)} \big]=R^{IJ}\big[\omega_{(LC)} \big]\ ,
    \label{curv3}
\end{equation}
while, the contorsion induced curvature must obey.
\begin{equation}
\begin{gathered}
    D_{\omega}K^{\prime IJ}-K^{\prime I}{}_{M}\wedge K^{\prime MJ}+
    D_{\omega}\theta^{IJ}+\theta^{IM}\wedge \theta^{J}{}_{M}=0 \ .
\end{gathered}
    \label{contcurv}
\end{equation}
Consider the on-shell integral of the 2nd term in (\ref{contcurv}) in the $e^{\prime I}=e^{\phi}e^{I}$ basis which is defined as
\begin{equation}
\label{Q0}
    \begin{aligned}
        Q_\theta[e,\omega_{(\mathrm{LC})},\phi]
        ={}&-\frac{1}{32\pi G}\int
        \epsilon_{IJKL}\,e^I\wedge e^J\,e^{2\phi}\\
        &\wedge\left[
            D_\omega\theta^{KL}
            +\theta^{KM}\wedge\theta^L{}_M
        \right].
    \end{aligned}
\end{equation}
Upon substituting the definition of $\theta^{IJ}$ in (\ref{omega02}) into (\ref{Q0}) $D_{\omega}\theta^{KL}$ is a boundary term such that the only contribution to (\ref{Q0}) is from $\theta^{KM}\wedge \theta^{L}_{M}$ which   can be expressed solely in terms of the conformal mode gradients. The integral  (\ref{Q0}) then takes the form
\begin{equation}
\label{Q2}
    \begin{aligned}
&Q_\theta[e,\omega_{(\mathrm{LC})},\phi]\\
        &\quad=-\frac{1}{32\pi G}\int
        \epsilon_{IJKL}\Bigl[
            e'^I\wedge e'^J\wedge e'^K\wedge e'^L
            \,(\nabla^M\phi\nabla_M\phi)\\
        &\qquad\quad
            -e'^I\wedge e'^J\wedge e'^K\wedge e'^M
            \,(\nabla_M\phi\nabla^L\phi)\\
        &\qquad\quad
            -e'^I\wedge e'^J\wedge e'^M\wedge e'^L
            \,(\nabla^K\phi\nabla_M\phi)
        \Bigr],
    \end{aligned}
\end{equation}
a rather atypical action for a scalar field. The integral $Q_{\theta}\big[e,\omega_{(LC)},\phi \big]$ in (\ref{Q2}) may be converted into metric variables using suitable $\epsilon-\delta$ identities for the rescaled tetrad $e^{\prime I}$ such that the $Q_{\theta}$ integral in metric variables is 
\begin{equation}
\label{Q3}
    Q_\theta[g_{\mu\nu},\phi]
    =-\frac{6}{16\pi G}\int d^4x\,\sqrt{g}\,
        e^{2\phi}(\nabla\phi)^2 ,
\end{equation}
the problematic kinetic term which renders the Euclidean Einstein-Hilbert action unbounded from below \cite{GHP1978}. Given the relation (\ref{contcurv}), it follows that $Q_{K}\big[e, \omega_{(LC)},\phi\big]$, the on-shell integral for the 1st term in (\ref{contcurv}),  obeys $Q_{K}\big[e, \omega_{(LC)},\phi\big]=-Q_{\theta}\big[e, \omega_{(LC)},\phi\big]$ so that the total contribution of (\ref{contcurv}) to the on-shell action vanishes. When expressed in metric variables, the integral $Q_{K}\big[e, \omega_{(LC)},\phi\big]$ is
\begin{equation}
    Q_K[g_{\mu\nu},\phi]
    =+\frac{6}{16\pi G}\int d^4x\,\sqrt{g}\,
        e^{2\phi}(\nabla\phi)^2 ,
\end{equation}
which cancels the problematic conformal kinetic integral so that the conformal mode problem is ultimately absent. As the conformal gradient term has been canceled on-shell,  evaluating (\ref{confEC}) on
(\ref{curveq}) yields the on-shell action
\begin{align}
    \begin{split}
        &S_{\mathrm{EC,E}}^{\mathrm{conf}}
        [e,\omega_{(\mathrm{LC})},\phi]=\\
        &-\frac{1}{32\pi G}\int
        \epsilon_{IJKL}\,e^I\wedge e^J
        \wedge \bigg[
            e^{2\phi}R^{KL}[\omega_{(\mathrm{LC})}]
            -e^{4\phi}\frac{\Lambda}{6}e^K\wedge e^L
        \bigg],
    \end{split}
\label{confEC2}
\end{align}
where the on-shell action obeys
\begin{equation*}
    S^{\rm conf}_{\rm EC,E}\big[e, \omega'_{(LC)} ,\phi\big]=S^{\rm conf}_{\rm EC,E}\big[e, \omega_{(LC)} ,\phi \big]
\end{equation*}
due to (\ref{curv3}), itself a consequence of remaining at fixed saddles $\omega^{IJ}_{(LC)}=\omega^{\prime IJ}_{(LC)}$. The cancellation of $Q_{K}\big[e, \omega_{(LC)},\phi\big]$ and $Q_{\theta}\big[e, \omega_{(LC)},\phi\big]$ ensures that the on-shell Einstein-Cartan action (\ref{confEC2}) is bounded from below along the conformal direction for $\Lambda>0$. In converting back to metric formulation, we find that the corresponding Euclidean Einstein-Hilbert action is
\begin{equation}
    S_{\rm EH,E(2)}^{\rm conf}[g, \phi]=\frac{-1}{16\pi G}\int d^{4}x\sqrt{g}\big[e^{2\phi}R-e^{4\phi} 2\Lambda\big]
    \label{confEH2}
\end{equation}
which is (\ref{confEH}) without the negative conformal kinetic term. This action is also conformally bounded from below for $\Lambda>0$. From this perspective, it's clear that the historic problem of (\ref{confEH}) being unbounded from below for all $\Lambda$ arises because torsion (or, contorsion) is automatically is equal to zero ($T^{I}=K^{I}{}_{J}\wedge e^{J}=0$) in general relativity. This in turn ensures that (\ref{conto02}) and (\ref{contcurv}), which cancel the conformal gradient and kinetic terms on-shell respectively, are impossible to implement. In Einstein-Cartan gravity, this is a choice rather than a necessity. However, this choice enforces (\ref{omega02}) so that saddles are no longer fixed and the usual instability is recovered. Hence, from the Einstein-Cartan perspective, the Einstein-Hilbert action with negative conformal kinetic term is not truly a saddle-point of the connection.

\begin{center}
    {\it Einstein-Hilbert $\rightarrow R^{2}/\Lambda$}
\end{center}
As in the Einstein-Cartan case, for $\Lambda>0$ the conformal saddle $y_{\rm min}=e^{2\phi_{\min}}$ of (\ref{confEH2}) can be found which is
\begin{equation}
    y_{\rm min}=e^{2\phi_{\min}}=\frac{R}{4\Lambda} \ .
    \label{phimin2}
\end{equation}
Once again, we see that $y_{\rm min}>0$ if both $R>0$ and $\Lambda>0$ such that classical de Sitter geometry is the conformal saddle of (\ref{confEH2}). Accordingly, the on-shell action evaluated on this saddle is
\begin{equation}
    S_{\rm EH,E(2)}^{\rm conf}[g, \phi_{\rm min}]=\frac{-1}{128 \pi G\Lambda}\int d^{4}x\sqrt{g}  \ R^{2}.
    \label{phishell2}
\end{equation}
The equations (\ref{phimin2}) and (\ref{phishell2}) are the Einstein-Hilbert equivalents of (\ref{phimin}) and (\ref{phishell}) up to numerical factors of $4/3$ and $3/2$ respectively. It follows that the saddle-point sequence
$S^{\rm conf}_{\rm EC, E}\big[e, \omega, \phi\big]\rightarrow S^{\rm conf}_{\rm EC, E}\big[e, \omega, \phi_{\rm min} \big]\rightarrow S^{\rm conf}_{\rm EC, E}\big[e, \omega_{(LC)}, \phi_{\rm min}\big]$ is equivalent to the sequence $S^{\rm conf}_{\rm EC, E}\big[e, \omega, \phi\big]\rightarrow S^{\rm conf}_{\rm EC, E}\big[e, \omega_{(LC)}, \phi \big]\rightarrow S^{\rm conf}_{\rm EC, E}\big[e, \omega_{(LC)}, \phi_{\rm min}\big]$ up to these factors. A consequence of this equivalence is that (\ref{phishell2}) possesses the same properties as (\ref{phishell}), namely, conformal invariance due to the transformations $\sqrt{g} \rightarrow e^{4\phi}\sqrt{g}$ and $R\rightarrow e^{-2\phi}R$ and a dimensionless coupling $1/G\Lambda$. Curiously, further substituting $R=4\Lambda$ into (\ref{phishell2}) reproduces the on-shell density $-\Lambda\sqrt{g}/8 \pi G$. so that the on-shell action is just de Sitter density $\times$ space-time volume.\\
The fact that (\ref{phishell2}) is a classical saddle of the conformally transformed Einstein-Hilbert (\ref{confEH2}) provides a rationale for a known result. Pure $R^2$ gravity is equivalent to Einstein gravity with a cosmological constant and minimally coupled scalar field, both in the metric theory \cite{AlvarezGaume2016} and with an independent connection \cite{EderyNakayama2019}. We reach that equivalence from the opposite direction. Starting from the Einstein-Cartan action, $R^2/\Lambda$ gravity is simply a conformal mode saddle point after connection saddle-point has been fixed as $\omega^{IJ}_{(LC)}=\omega^{\prime IJ}_{(LC)}$ by (\ref{conto02}).
\begin{center}
    {\it Discussion \& Conclusion}
\end{center}
%With boundedness of the Einstein-Cartan and Einstein-Hilbert actions for $\Lambda>0$ demonstrated technically, we now turn our attention to more conceptual matters. Specifically, the philosophical implications of this boundedness. 
Ultimately, the conformal instability and the gravitational path integral are inseparable as the problem was first formulated in this context. Given the resolution we've proposed, the correct starting point for path integral analysis appears to be the quantum gravitational path integral \cite{Arkady, Sue, Kondo, IM}
\begin{equation}
    Z_{\rm EC}=\int {\cal D}e\,{\cal D}\omega\,
\exp\left[-{1\over\hbar}S_{\rm EC,E}[e,\omega]\right]\ ,
\label{Z}
\end{equation}
defined by the action (\ref{ECE}) \cite{N3}. Within this path integral ${\rm exp}\big[-S_{\rm EC,E}[e, \omega]/\hbar  \big]$ is a statistical weight which assigns probabilities to tetrad and connection configurations. The path integral (\ref{Z}) therefore sums over equilibrium connection configurations defined by (\ref{torsioneq}) and equilibrium tetrad configurations defined by solutions to
\begin{equation}
    e^{J} \wedge \bigg[R^{KL}[\omega]-\frac{\Lambda}{3}e^{K}\wedge e^{L}\bigg]=0 \ ,
\end{equation}
as well as statistically disfavoured configurations with respect to both fields. Yet, with respect to conformal transformations, the statistical weight  ${\rm exp}\big[-S_{\rm EC,E}[e, \omega]/\hbar  \big]$ is only meaningful (i.e. conformally bounded from above by ${\rm exp}\big[-S_{\rm EC,E}[e, \omega, \phi_{\rm min}]/\hbar  \big]$) when $\Lambda>0$. Could it be, that the statistical distribution ${\rm exp}\big[-S_{\rm EC,E}[e, \omega]/\hbar  \big]$ diverging for $\Lambda\leq 0$ precludes these geometries from existing? Conversely, could the fact that it's well behaved for $\Lambda>0$ indicate that only these geometries may be physically realised? This would offer a particularly tantalising rationalisation for why our universe is $\Lambda>0$ as opposed to the alternatives.\\
That being said, arguments of this nature should be made with caution. An obvious rebuttal to this heuristic is that the same argument applied to the canonical Einstein-Hilbert statistical weight ${\rm exp}\big[-S_{\rm EH,E}[e, \omega]/\hbar  \big]$ would render all geometries unphysical. Yet, from this same weight, the equilibrium partition function for the static patch of $\Lambda>0$ is derived. Viewing the Einstein-Hilbert action as the connection saddle of the Einstein-Cartan action for fixed $\omega^{IJ}=\omega^{IJ}_{(LC)}=\omega^{\prime IJ}_{(LC)}$ serves as vindication for why the static $\Lambda>0$ partition function calculation was historically taken to be true in spite of the conformal factor problem.\\
To this end, we note that the equilibrium partition function for a Schwarzschild black hole (where $\Lambda=0)$, among other examples is also derived from ${\rm exp}\big[-S_{\rm EH,E}[e, \omega]/\hbar  \big]$. These results are widely believed to hold deep truths about the ``quantum" nature of space-time in spite of the conformal factor problem (along with other outstanding issues). Applying the same rationale as the $\Lambda>0$ case seems to point to an intriguing possibility. Namely, that Einstein-Cartan gravity (and standard general relativity by association) could be considered as classical saddles of a more general formulation of gravity which completely resolves the conformal mode problem for all $\Lambda$.\\

\paragraph{Acknowledgments} RI would like to thank Michael Fine for enlightening discussions. SA is supported by the Simons Foundation Award 896696. RI is supported by the Institute of Physics through the Bell Burnell Graduate Scholarship Fund.  JM is partly supported by STFC Consolidated Grant ST/X000575/1 and Consolidated Award UKRI4230


\begin{thebibliography}{99}
\bibitem{N1}
These conformal transformations are not unique and it is possible to consider transformations for which $\omega^{IJ} \rightarrow f(\phi) \ \omega^{IJ}$, where $f(\phi)$ is an arbitrary function of the conformal mode. Nonetheless, the transformations (\ref{weyl}) are chosen since they're the simplest possible and return the Einstein-Hilbert action and the historic conformal kinetic action on-shell.
\end{thebibliography}
\end{document}